\documentclass[11pt,twoside]{article}
\usepackage{asp2010}

\resetcounters

\begin{document}

\title{Challenges in Solar and Stellar Model Physics}
\author{J.~A.~Guzik
\affil{XTD-NTA, MS T-086, Los Alamos National Laboratory, Los Alamos, NM 87545-2345 USA}}

\begin{abstract}

We are reaching relative maturity and standardization in one-dimensional single-star stellar evolution and pulsation modeling, and are making advances in binary and 2D and 3D models.  However, many physical inputs are still uncertain or neglected in models of the Sun and of other stars.  Thanks to the {\it Kepler}, CoRoT, and MOST spacecraft, and ground-based networks, we now have pulsation data for stars that are of comparable quality to that for the Sun to constrain models and test physical assumptions.  Here I will focus on main sequence (core H-burning) or slightly post-main sequence (shell H-burning) stellar models, and some of the unsolved problems for these stars.  I will revisit the solar abundance problem, and show the effects of modified electron screening,  dark matter, and early mass loss on solar models.  I will discuss the $\gamma$ Dor/$\delta$ Sct hybrid stars, the mismatch between predicted and observed frequencies for $\delta$ Sct stars,  and how seismology of stars more massive than the Sun, e.g. the brightest {\it Kepler} target $\theta$ Cyg, could help us constrain physical processes such as diffusive settling, test pulsation driving mechanisms, and provide clues to the solar abundance problem.  
\end{abstract}

\section{Introduction}

In recent years we have found that stars of nearly every evolutionary stage exhibit pulsations \citep{2010aste.book.....A}.  Some outstanding problems for main sequence (core hydrogen-burning) or slightly post-main sequence (shell hydrogen-burning) stars include the solar abundance problem \citep[see, e.g.,][]{2008PhR...457..217B,2010ApJ...713.1108G}, the origin of low metallicity in $\lambda$ Boo stars \citep[see, e.g.,][]{2010MNRAS.406..566M, 2002ApJ...573L.129T}, discrepancies between predicted and observed frequencies in evolved $\delta$ Sct stars \citep[e.g.,][]{2000ASPC..210..247G, 2004ASPC..310..462G}, hybrid $\gamma$ Dor/$\delta$ Sct stars \citep[e.g.,][]{2011A&A...534A.125U}, and $\beta$ Cep/SPB stars (e.g., \citep[e.g.,][]{2008MNRAS.385.2061D}.  See also \citet{2013ASSP...31..201G} and references therein for additional discussion of some of these issues.

\section{Challenges and Constraints}

Stellar evolution and pulsation modeling requires choices for physics approximations, databases, and numerical methods.  Observations are used to test the adequacy and validity of models and approximations.  Below are listed considerations in each category; most are still the subject of ongoing research.

\subsection{Physics challenges}

\begin{itemize}

\item Convection modeling (mixing length theory, or a more advanced theory, treatment of semi-convection and convective overshooting)
\item Pulsation-convection interactions (time-dependence)
\item Element diffusion (gravitational, thermal, chemical) and radiative levitation
\item Mixing processes (meridional circulation, waves, shear)
\item Rotation (differential, angular momentum transport)
\item Mass loss or accretion
\item Magnetic fields
\item Binarity, tidal effects
\item Dark matter
\item Pulsations, oscillations, waves (gravity, acoustic)
\item Mode coupling, mode amplitudes (nonlinear effects)
\item Viscosity (turbulent, molecular)
\item Pulsation driving and mechanisms ($\kappa$ effect, stochastic excitation, convective blocking, convective driving, epsilon mechanism)
\item Adiabatic vs. nonadiabatic effects
\item Turbulent pressure and energy
\end{itemize}

Most evolution models for main-sequence stars use the mixing-length theory of convection, and neglect mass loss (unless M $>$ 20 M$_{\odot}$), accretion, diffusion, and magnetic fields.  Some codes, for example YREC \citep{2008Ap&SS.316...31D}, include the effects of rotation, angular momentum transport, and turbulence in a parametrized form.  Many physical processes have been explored individually in stellar evolution or pulsation models to explain particular observations, for example instability region boundaries, chemical peculiarities, or rotational splitting of modes.    The accurate and extensive observational data on the Sun, our closest star, provide the most rigorous tests and constraints for some of these processes.

\subsection{Stellar model input and databases}

\begin{itemize}
\item Abundances and abundance mixtures
\item Opacities (radiative, conductive, monochromatic, molecular)
\item Equation of state (relativistic effects, electron exchange, ionization, excitation)
\item Diffusion coefficients
\item Nuclear reaction rates (electron screening effects)
\item Fundamental constants (gravitational constant G, Stefan-Boltzmann constant $\sigma$, solar mass, solar radius, etc. )
\item Stellar atmosphere models
\end{itemize}

Stellar models require the inputs above to provide initial and boundary conditions  to solve the equations of stellar structure and evolution, and to include additional processes such as diffusive settling.  Opacities are usually included by interpolating in pre-calculated tables, e.g. OPAL \citep{1996ApJ...464..943I}, OP \citep{1995AIPC..322..117S}, or LEDCOP \citep{2001ApJ...561..450N}, supplemented by conductive opacities \citep[e.g.,][]{2007ApJ...661.1094C}, and low-temperature opacities \citep[e.g.,][]{2005ApJ...623..585F}.  The equation of state is included either using tables, e.g. OPAL \citep{1996ApJ...456..902R}, MHD \citep[see, e.g.,][]{2006ApJ...646..560T}, or analytical forms, e.g., CEFF \citep{1992A&ARv...4..267C}, the Irwin Free-EOS \citep{2012ascl.soft11002I}, MHD or OPAL emulators \citep{2010Ap&SS.328..175L}, or  SIREFF \citep{1997ApJ...491..967G}.  Coefficients for thermal diffusion \citep{1986ApJS...61..177P} or monochromatic opacities for radiative levitation \citep{1998ApJ...492..833R} are required to treat these processes.  Nuclear reaction rates continue  to be updated \citep{2011RvMP...83..195A}.  Many stellar evolution codes were written in the early 1960s, and the physical constants or solar parameters adopted then may have undergone significant revision.  Some constants or dimensions may be combined as a coefficient in front of an expression, so it is worthwhile to check one's code for consistency and use of up-to-date values.

\subsection{Numerical challenges}

\begin{itemize}
\item Disparate time and spatial scales for different processes
\item Zoning (1D, 2D, 3D, rezoning, adaptive grid)
\item Eulerian (fixed spatial grid) vs. Lagrangian (grid follows mass motion)
\item Subgrid models
\item Interpolation, extrapolation in tabular inputs
\item Timestep control (implicit, explicit)
\item Boundary conditions (surface, center)
\item Linear vs. nonlinear
\item Viscosity treatment
\end{itemize}

Because physical processes happen on disparate time and spatial scales, specialized codes with appropriate assumptions are needed to isolate or study a given process.  One-dimensional standard stellar evolution models may include several hundred radial zones and several hundred timesteps to follow evolutionary changes and nucleosynthesis.  Processes such as rotational mixing or convective overshooting are sometimes treated in a parametrized manner based on simulations from 2D or 3D hydrodynamic codes with smaller time steps and zone size.  The smallest turbulent length scales cannot be resolved, and subgrid models are used to approximate processes happening at these scales.  Many pulsation models use a linear perturbation theory to find the normal mode frequencies of a static model, while nonlinear hydro codes are used to model large-amplitude pulsations as found in $\delta$ Sct, $\beta$ Cep, Cepheid or RR Lyr variables.  The pulsation amplitudes are affected by viscosity treatments \citep{2007MNRAS.377..645S}.  Viscosity is  likely also important in damping nonradial modes with a lot of horizontal shear \citep{2000ApJ...542L..57G}. 

\subsection{Observational tests and constraints}
\begin{itemize}
\item Photometry (multicolor)
\item Spectroscopy (surface abundances and gravity, temperature, B fields, mass loss)
\item Binaries (eclipsing)
\item Clusters
\item Magnetic fields, magnetic field-induced activity (spots, flares)
\item Pulsation spectrum, amplitudes, phases, variations with time
\item Neutrinos
\end{itemize}

Inferences from stellar pulsations have over the last 50 years allowed us to see inside a star; neutrino observations of the Sun have provided constraints on its nuclear burning regions.  Eclipsing binaries and stars in clusters having a common age and initial element abundance provide very useful constraints on stellar evolution.  Explaining how features observed at the surface are produced has led to hypotheses that can be checked for consistency by examining similar behavior of a larger set of stars.

\section{Example challenges}

A few physics challenges to be discussed next include: The solar abundance problem, the prevalence of hybrid $\gamma$ Dor/$\delta$ Sct (and SPB/$\beta$ Cep) stars; and the mismatch between pulsation theory prediction and observation for evolved $\delta$ Sct stars.   A few other problems for main-sequence stars that will not be discussed here are:  The origin of low metallicity in $\lambda$ Boo stars, and apparently non-pulsating stars in pulsation instability regions \citep [see, e.g.,][]{GuzikConstantStar2013}.
  
\subsection{Solar abundance problem}

\articlefiguretwo{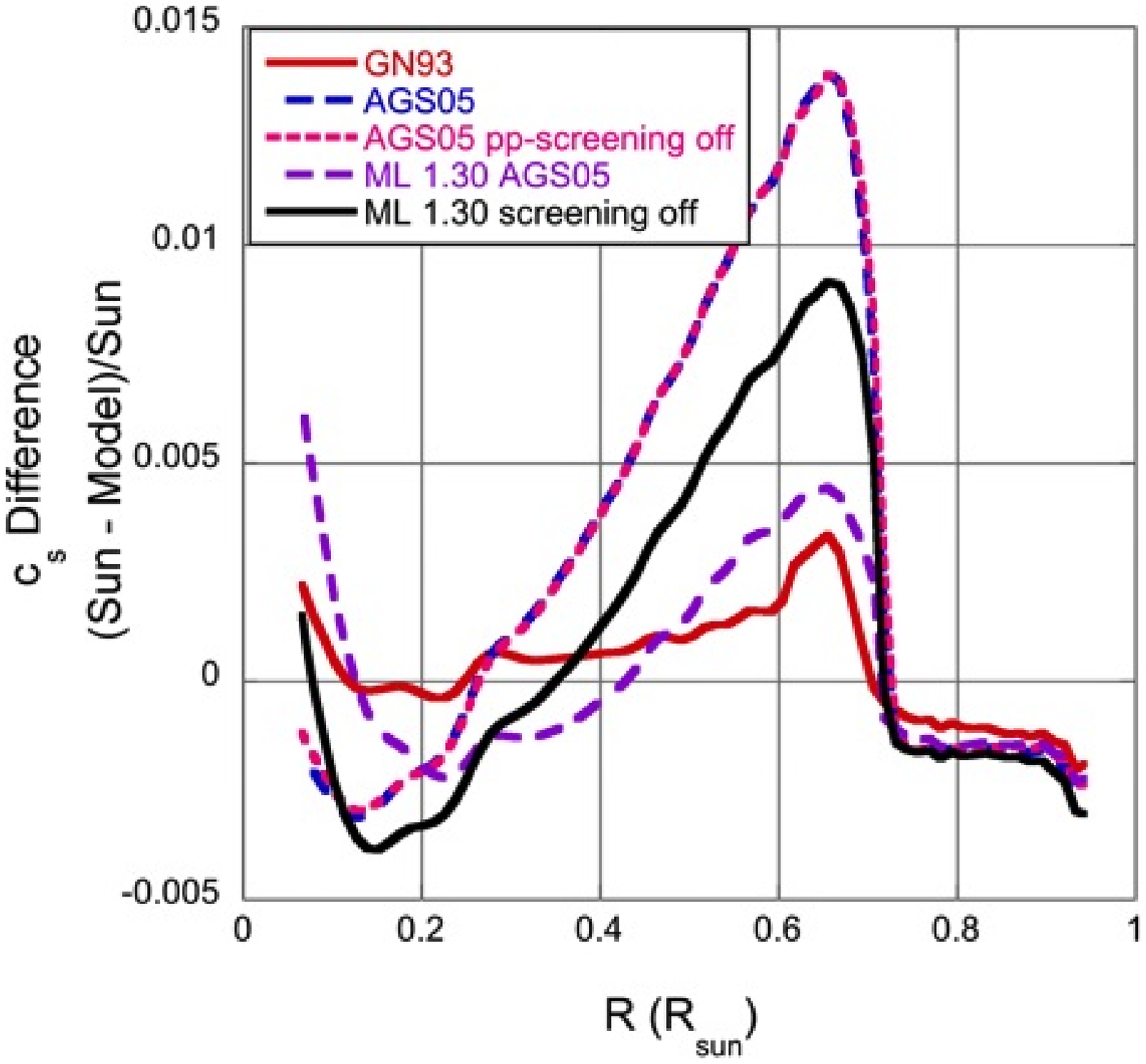}{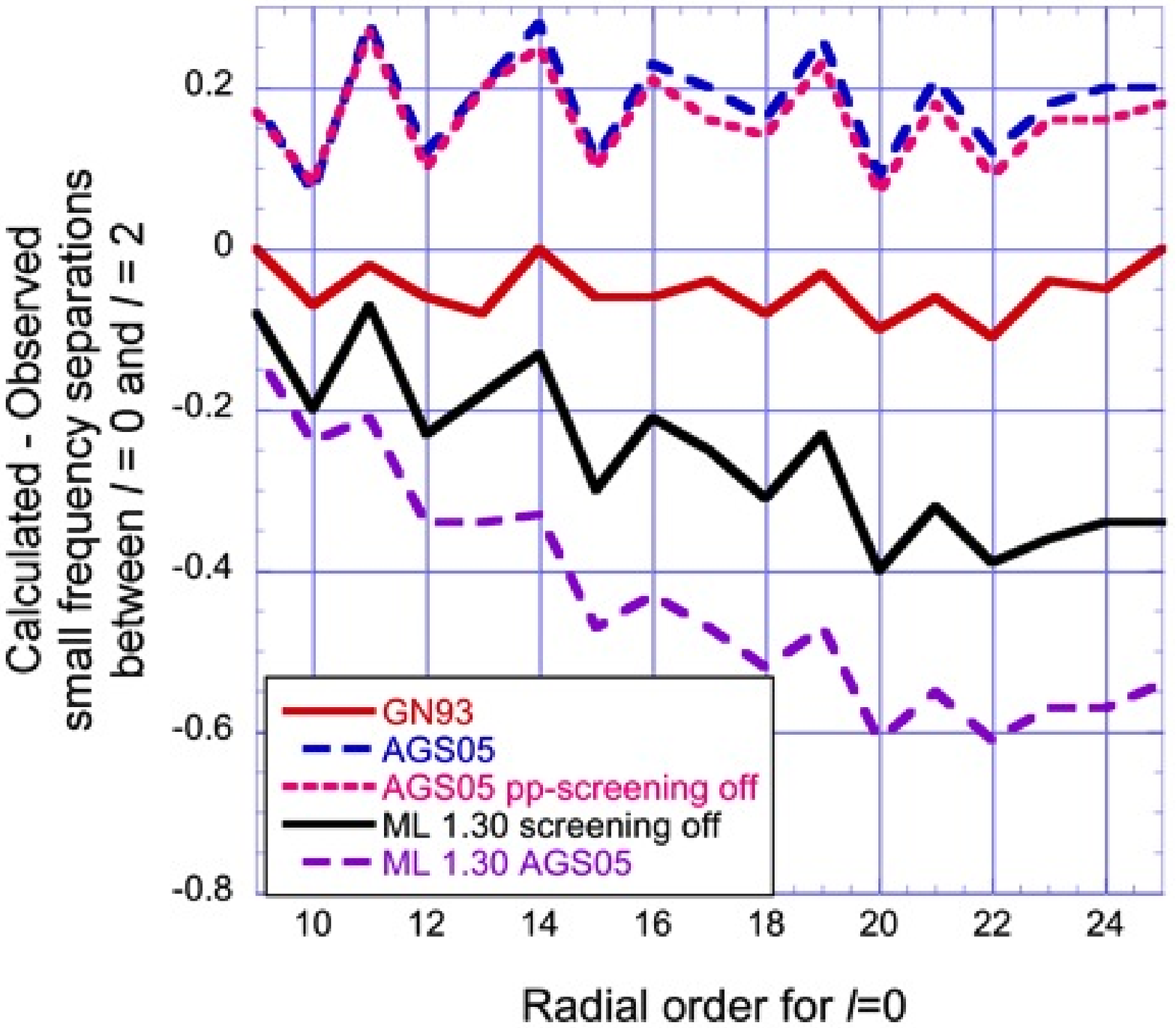}{screening}{Left:  Difference between helioseismically inferred sound speed vs. radius for models with GN93 and AGS05 solar abundances, with and without early mass loss of 0.3 M$_{\odot}$, and with and without electron screening in reaction rates.  The helioseismically inferred sound speed is from \citet{2000ApJ...529.1084B}. Early mass loss can significantly improve sound speed agreement with inferences for the lower AGS05 abundances; the effect of turning off electron screening in the p--p reactions, plus reducing screening in the CNO-cycle reactions of the more massive early mass-loss model is large.  Right:  Calculated minus observed small frequency separations (differences  between $\ell$ = 0 and $\ell$ = 2 mode frequencies separated by one radial order).  Turning off screening for the mass-losing model improves the agreement with observations.}

Since the abundances of elements heavier than H and He were revised downward by \citet{2005ASPC..336...25A} (hereafter AGS05), solar evolution models using these abundances no longer show the relatively good agreement between stellar models and helioseismic constraints seen with the older abundances, e.g., \citet{1993oee..conf...15G} (GN93), or \citet{1998SSRv...85..161G}.   Many changes to standard solar models have been explored to improve the agreement \citep[see, e.g.,][]{2008PhR...457..217B, 2010ApJ...713.1108G}, including modified opacities, enhanced diffusion, mass loss, accretion, and increased Ne abundance.  Abundance determinations are continuing to be reassessed, and have increased somewhat since 2005 \citep[see, e.g.,][]{2009ARA&A..47..481A,2011SoPh..268..255C,2011JPhCS.271a2034A}, making the problem less severe.  The process of dealing with the solar abundance problem has caused modelers to re-examine assumptions and investigate modifications to the standard solar model.  Below we discuss the effects of early mass loss, electron screening of nuclear reaction rates, and dark matter.

Increasing the mass and luminosity of the Sun at the beginning of its lifetime has advantages for explaining early warmer temperatures on Earth, and depletion of solar surface lithium by about a factor of 200  from its initial value.  However,  if the Sun remained at higher mass for too long, the now-surface layers would have experienced temperatures early in the Sun's lifetime that would have destroyed all of the lithium.  Therefore, the observation of some surface Li constrains the amount and rate of mass loss \citep[see][]{2010ApJ...713.1108G, Wood2014}.

Molecular dynamics (MD) simulations show that the Salpeter screening formulas used a basis for screening nuclear reactions in stellar models may overestimate the screening \citep{2006ESASP.624E..20M,2009ApJ...701.1204M,2011ApJ...729...96M}.  For the proton-proton reaction, static screening reduces the reaction rate by about 5\%\ in the Sun's core.  According to the MD simulations, since most nuclear reactions occur between fast-moving ions at the high-energy tail of the Maxwellian velocity distribution, these fast moving nucleons feel the screening effects of the electrons less, and the reaction rate turns out to be nearly the same as for unscreened reactions.

\citet{Wood2014} are exploring the implications of reduced screening for helioseismology and the solar abundance problem.  Figure~\ref{screening} (left) shows the difference between helioseismically inferred (Basu et al. 2000) and calculated sound speed profile for solar models with the GN93 or AGS05 abundances, and with or without 0.3 M$_{\odot}$ of early mass loss, and with or without electron screening on the proton-proton reaction.  For the mass-losing model, screening was also reduced in the C+p and N+p reactions of the CNO-cycle.  Early mass loss improves the sound speed agreement for models using the AGS05 abundances.  For the model with initial mass 1.3 M$_{\odot}$ and AGS05 abundances, reduced screening has a larger effect due to the effect on the dominant CNO-cycle reactions early in the model evolution.  Figure~\ref{screening} (right) shows the differences between calculated and observed small frequency separations (differences between $\ell$=0 and $\ell$=2 mode frequencies separated by one radial order).  The small frequency separations are a more sensitive test of the solar core structure than are the sound speed inferences, and show the best agreement for the standard no-mass loss model with GN93 abundances.  Turning off/reducing screening in the p--p/CNO-cycle reactions for the mass-losing model improves agreement with observations for the small separations.

An initial solar mass of 1.3 M$_{\odot}$ would probably result in destruction of all present-day surface  Li; also, molecular dynamics simulations aren't available yet to quantify reduced screening for CNO-cycle reactions.  However, Fig. \ref{screening} shows that early mass loss and screening deserve further consideration in conjunction with abundances for calculating solar and stellar models and comparing with seismic inferences.

\articlefigure[width = 7.5 cm]{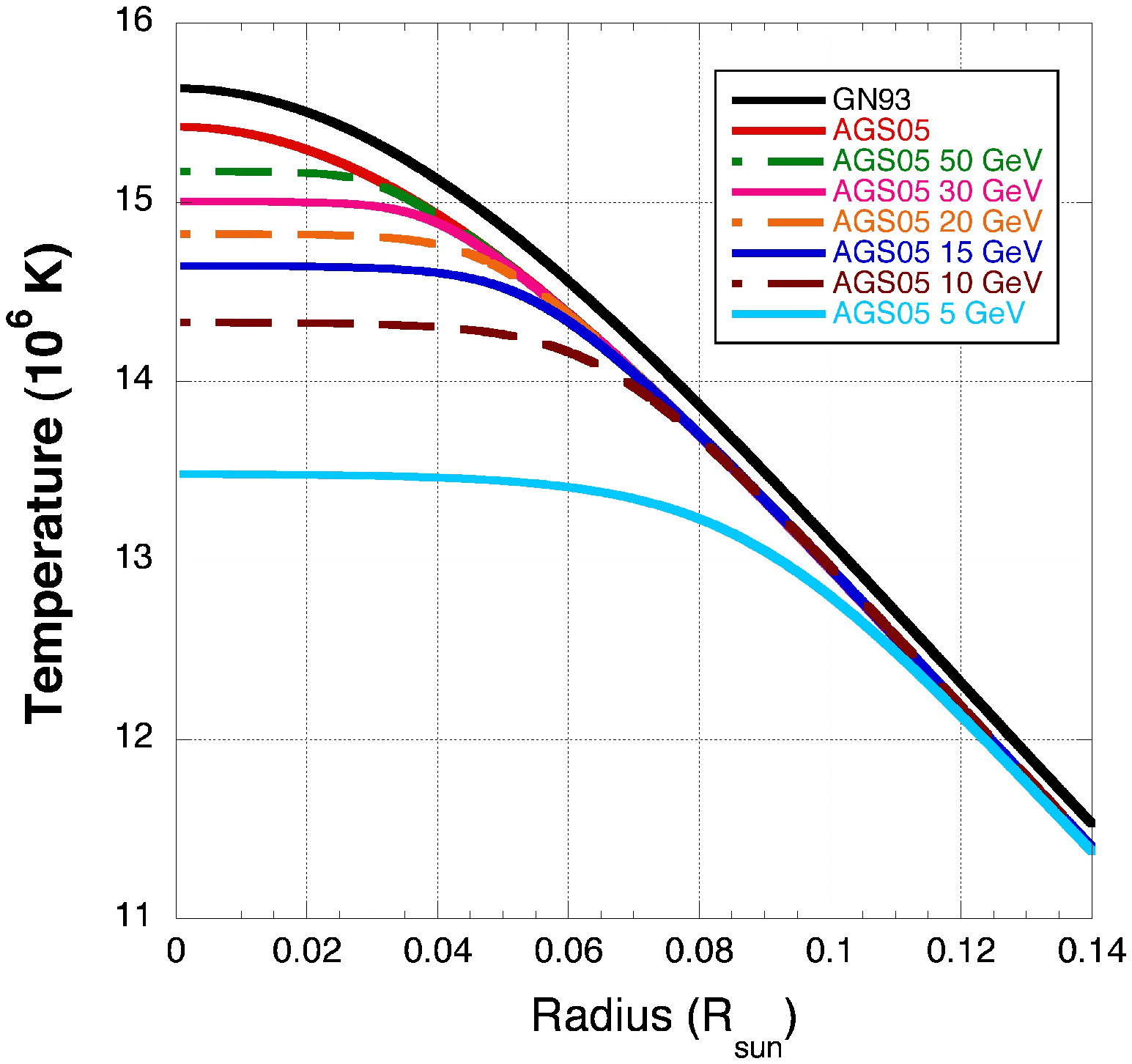}{centraltempWIMP}{Temperature vs. radius in core of evolved solar models with and without WIMP energy transport for WIMPs of various masses in GeV/c$^2$.  WIMPs transport energy from the inner to outer core, and reduce the central temperature.}

\articlefiguretwo{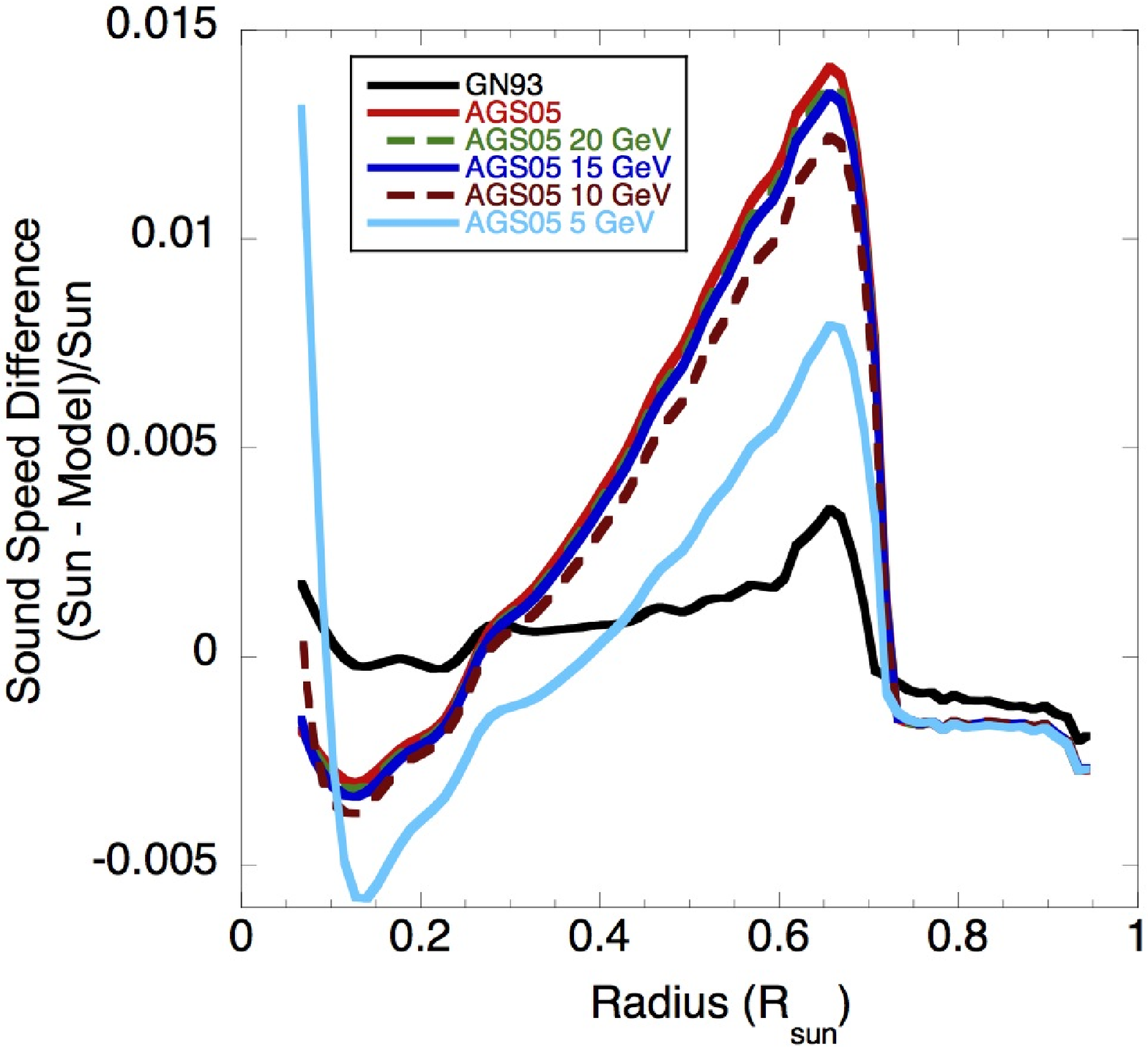}{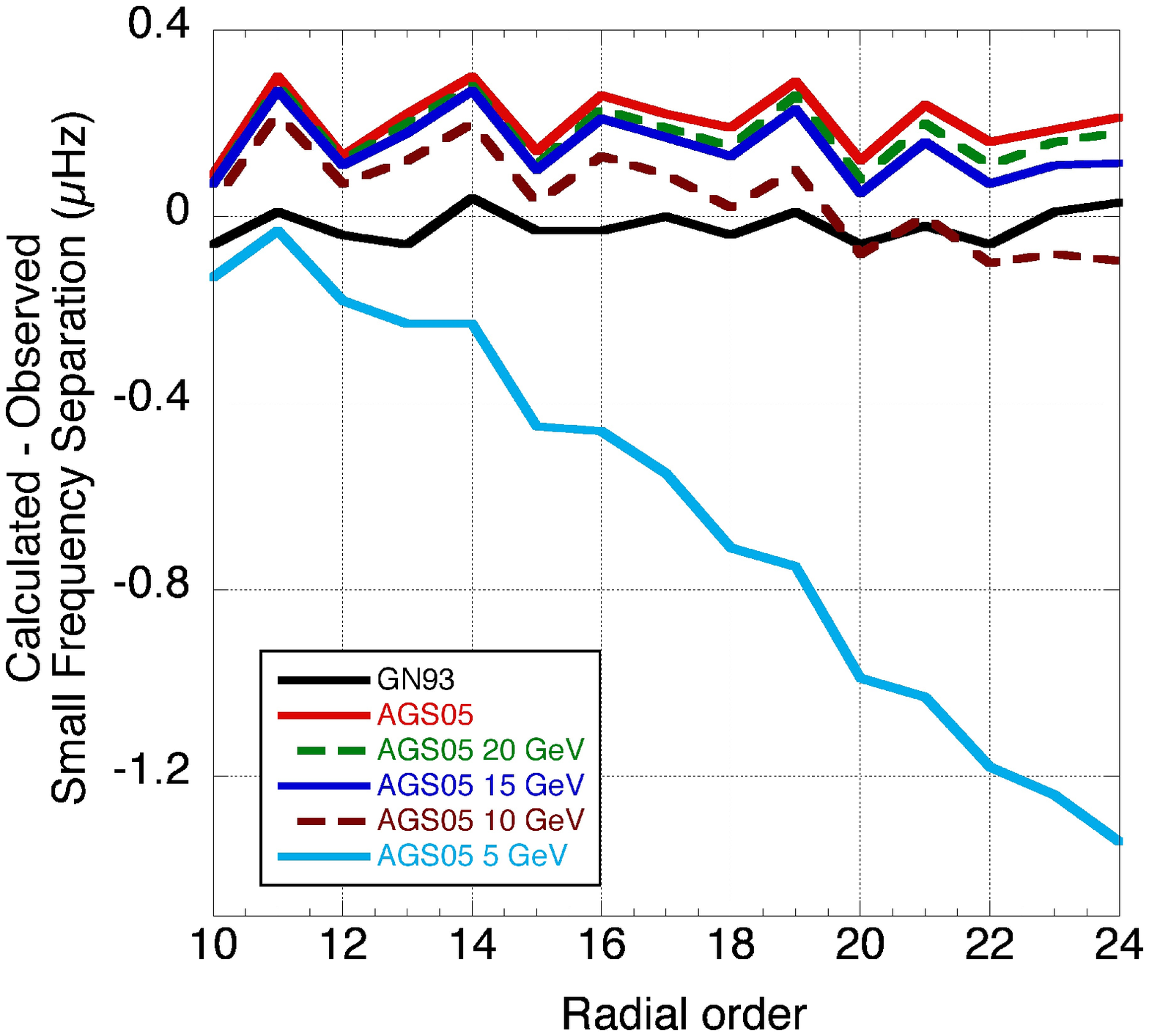}{WIMPs}{Sound speed (left) and small frequency separation (right) differences between models and helioseismic inferences for solar models with and without WIMP energy transport, for WIMPs of various masses.  Helioseismology probably rules out 5 GeV/c$^2$ WIMPs for the large interaction cross section used in these models.}

Dark matter, in the form of weakly interacting massive particles (WIMPs), is thought to make up 25\% of the mass of the universe.  Dark matter detection experiments are narrowing parameter space for the mass and interaction cross sections of dark matter particle candidates \citep{2010ARA&A..48..495F}.    WIMPS orbiting through the center of the Sun weakly interacting with matter would transport energy from the inner to outer core, producing an isothermal, cooler core, and would also reduce the output of pp-II and pp-III chain neutrinos.   In stellar models, WIMPs can be treated as a heat conduction process by modifying the opacity profile.  The effects of WIMPs have been included in solar and stellar models to see whether observations place additional constraints on WIMP properties (\citep{2010PhRvD..82j3503C,2012ApJ...746L..12T,2012ASPC..462..537H,2013arXiv1307.6519C}).  Figure~\ref{centraltempWIMP} shows the effect on solar core temperature, and Figure~\ref{WIMPs} shows the sound speed and small frequency separation comparisons for standard solar models and models including WIMPs with an interaction cross section of 7 $\times$ 10$^{-35}$ GeV/cm$^2$ \citep[see also][]{2010PhRvD..82j3503C}. While an interaction cross-section as large as used for these models is now ruled out by WIMP detection experiments, light WIMPs of 5 -- 7 GeV/cm$^2$ are not ruled out.  WIMPs may have a larger effect for stars near galactic centers where the WIMP density is larger \citep{2012PhRvL.108f1301I}, or in stars with more centrally condensed cores where WIMPs can not evaporate as rapidly and interact more frequently.  \citet{2013arXiv1307.6519C} find that WIMP energy transport may suppress convective cores in stars slightly more massive than the Sun; asteroseismology of stars such as $\alpha$ Cen B may place more stringent constraints than helioseismology on WIMP properties. 

\subsection{Evolved $\delta$ Sct stars}

\articlefigure[width=6.0 cm]{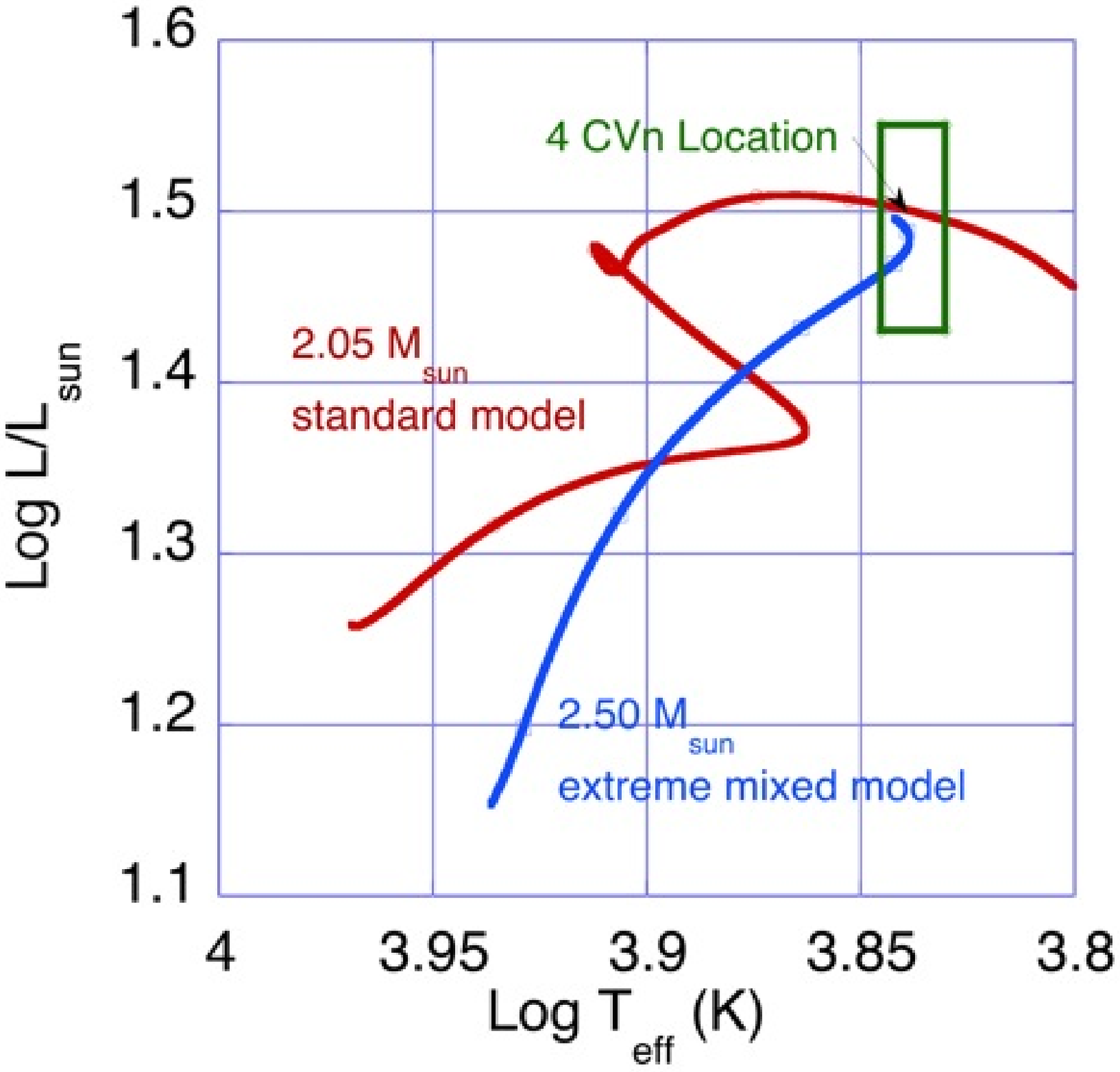}{4CVnHRD}{H -- R diagram showing evolution tracks of standard and mixed-core 4 CVn models.}

\articlefigure[width=9.0cm]{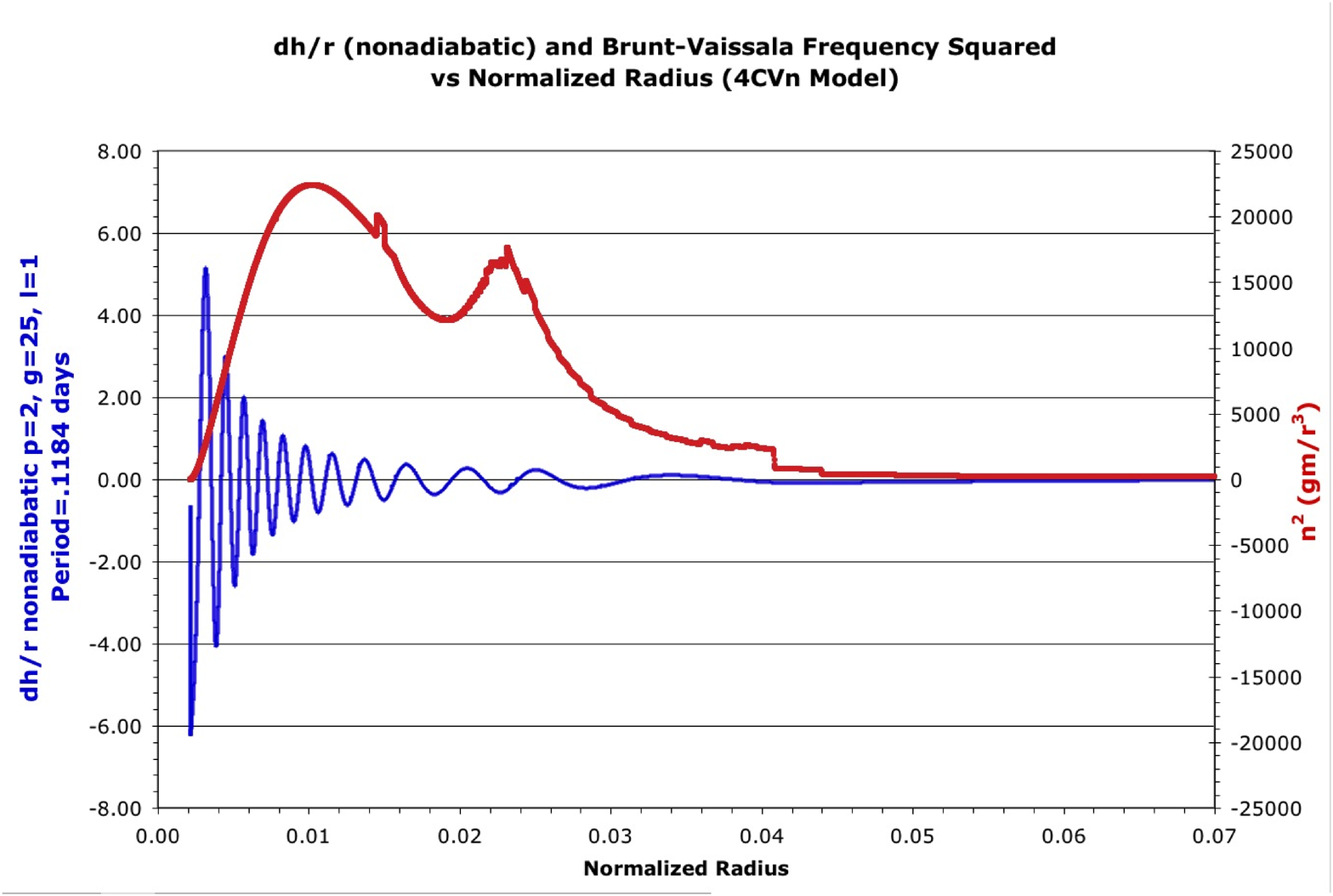}{4CVnBV}{Example horizontal {\it g}-mode eigenfunction and Brunt-V\"ais\"al\"a frequency in core of 4 CVn model.  The {\it g}-mode cavity that develops after core hydrogen is exhausted and the convective core disappears results in a dense frequency spectrum of mixed modes to be predicted.}

\articlefiguretwo{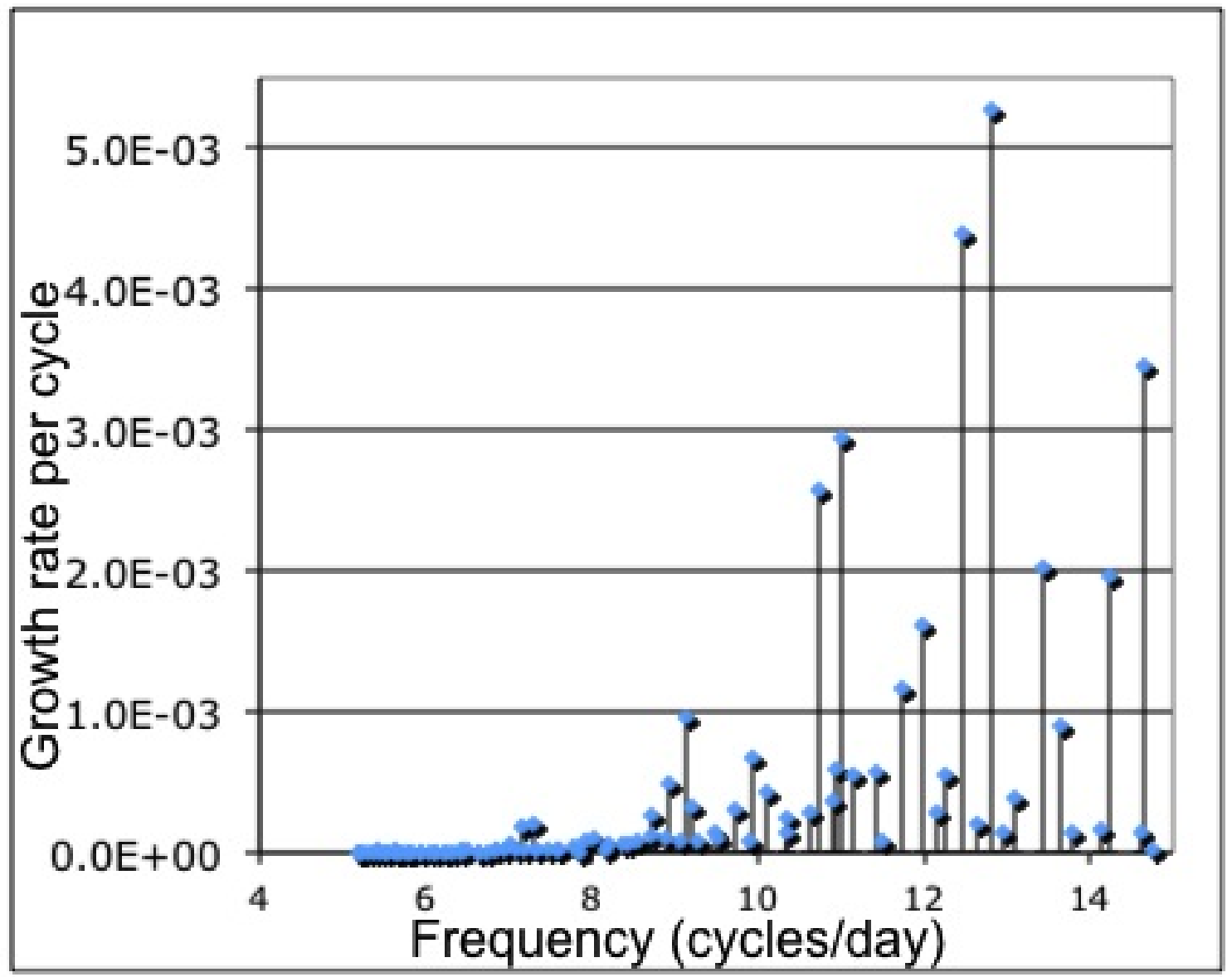}{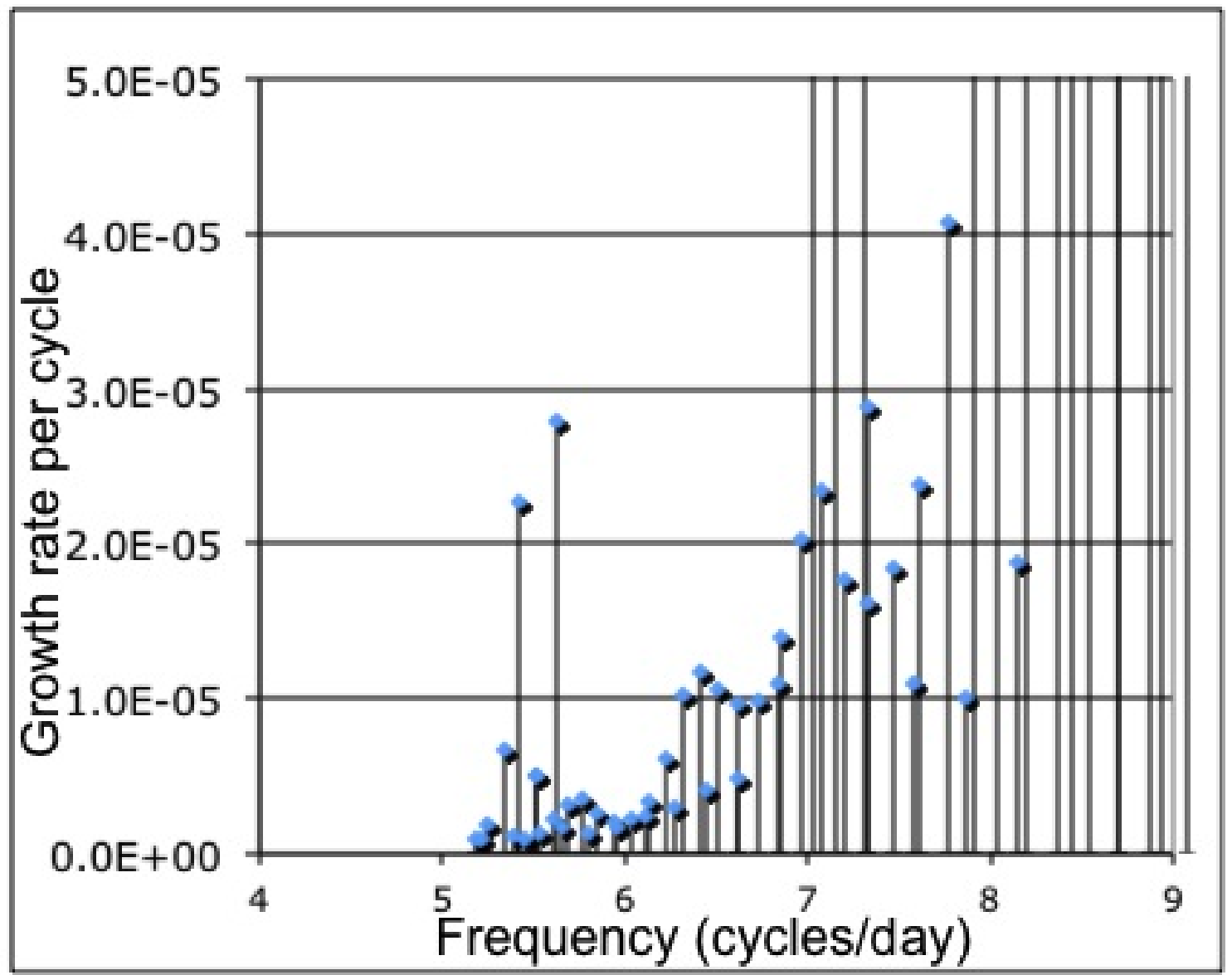}{4CVnfreqs}{Calculated unstable $\ell$ = 0, 1, and 2 modes and growth rates for standard evolution model of 4 CVn; rotational splitting is not included, and would further increase the density of frequencies in this plot.  The plot on the right has an enlarged scale to better show the lower-frequency, lower growth-rate modes.}

For some bright $\delta$ Sct stars with extensive photometric and spectroscopic observations, the spectrum of $\ell$ = 0, 1, and 2 modes that is predicted by pulsation models and should be detectable is not observed.  These stars do not have high-order modes that would have near-uniform frequency spacings; mode coupling, selection, and possibly rapid and differential rotation also make mode identification difficult.  For the evolved $\delta$ Sct star 4 CVn, ground based multisite photometric campaigns found 18 independent modes \citep{1999A&A...349..225B}.  Photometric and spectroscopic observations were used to identify 12 of the highest-amplitude modes \citep{2008CoAst.157..124C}.  We calculated stellar models with luminosity and T$_{\rm eff}$ appropriate for 4 CVn.  Observational constraints on luminosity and T$_{\rm eff}$ put this star in a region of the H -- R diagram where, according to standard stellar evolution models, it has finished core H-burning and has transitioned to shell H-burning (Fig. \ref{4CVnHRD}).  Since the model no longer has a convective core, the core has a non-zero Brunt-V\"ais\"al\"a frequency where nonradial gravity modes with a large number of {\it g}-type nodes can exist (Figure\ref{4CVnBV}).  These modes are actually mixed modes, and have a {\it p}-mode character in the stellar envelope.  For our models with no convective overshooting, our nonadabatic code predicts 332 $ell$ = 0, 1, 2 closely-spaced modes to be pulsationally unstable in the observed frequency range (Figure\ref{4CVnfreqs}.)

To attempt to reduce the number of predicted modes, we calculated a model with artificially enhanced core mixing in order to extend the core-hydrogen-burning convective-core phase (Figure\ref{4CVnHRD}).  This model has nearly twice the age of the standard model (1.6 Gyr instead of 0.9 Gyr) when it reaches the location of 4 CVn in the H -- R diagram.  We were able to reduce the number of predicted modes with this model to 30 instead of 332; however, the mode frequencies and spacings do not correspond well to those observed in 4 CVn.  See also \citet{2000ASPC..210..247G,2004ASPC..310..462G} and references therein for additional discussion.

\subsection{Hybrid $\gamma$ Dor and $\delta$ Sct stars}

\articlefiguretwo{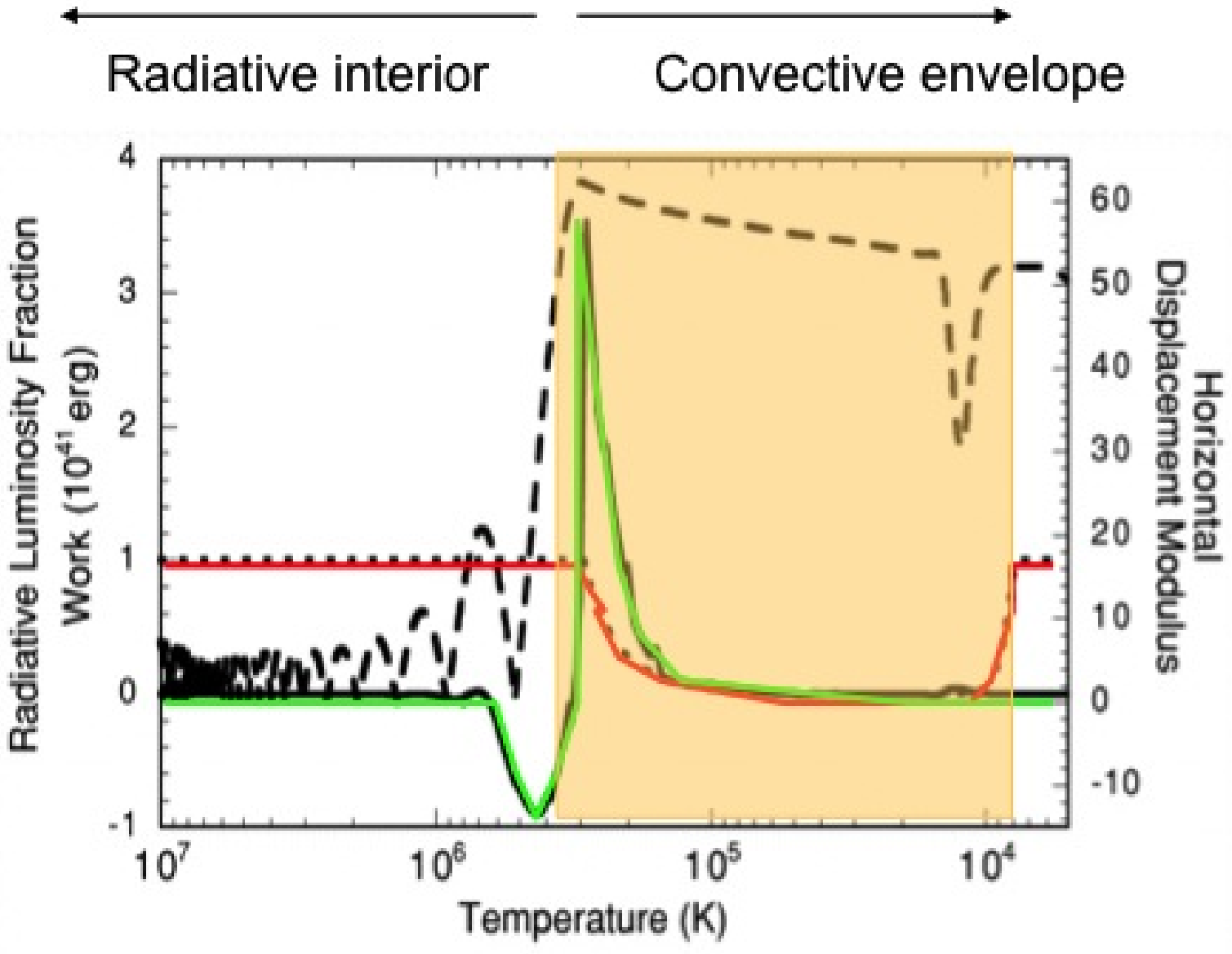}{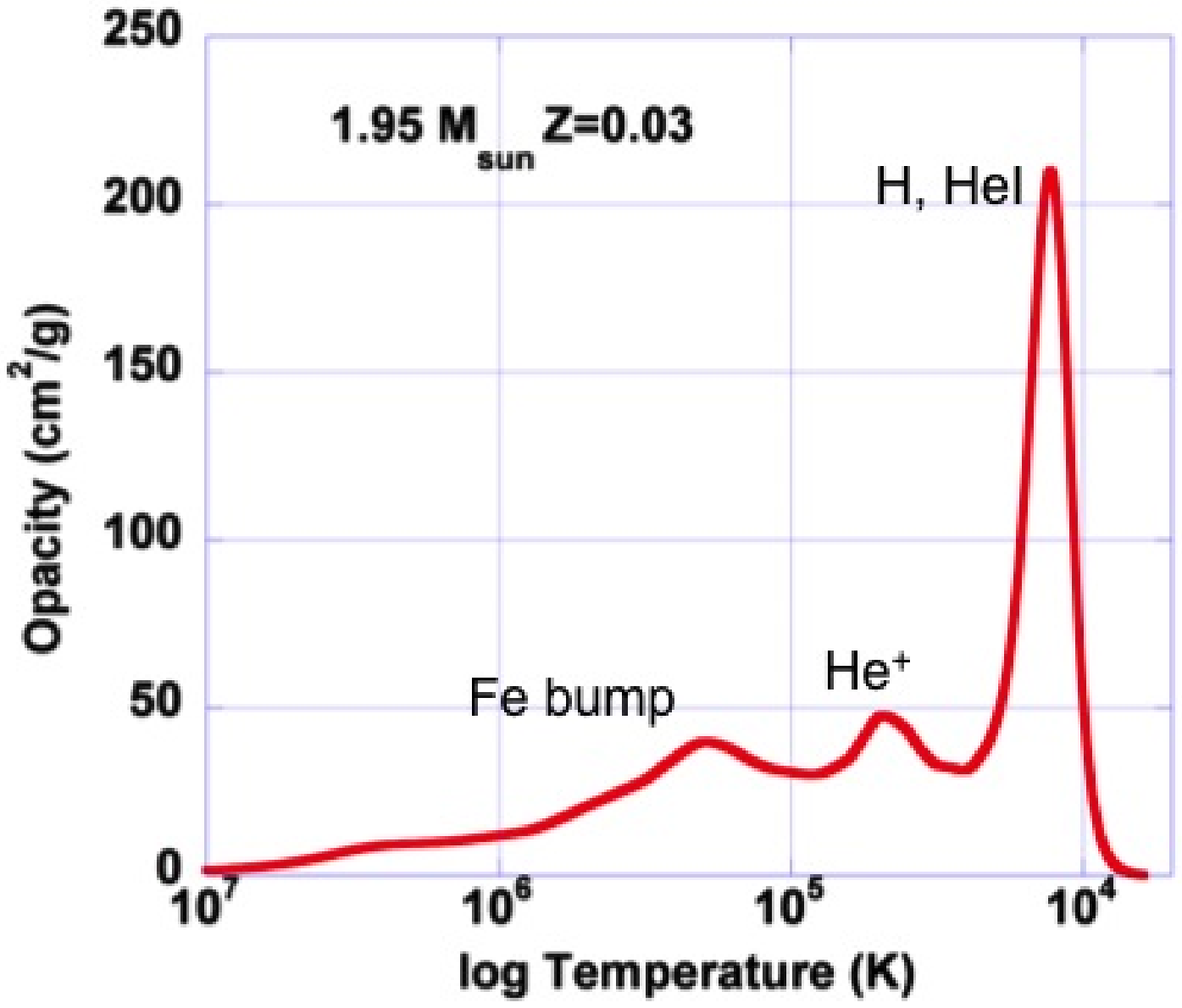}{Driving}{Left: Driving region below base of convection zone for a $\gamma$ Dor model \citep{2000ApJ...542L..57G}; Right: Opacity vs. temperature for a 1.95 M$_{\odot}$ Z=0.03 model showing bumps that modulate radiation flow.  The He+  ionization bump near 50,000 K is responsible for driving $\delta$ Sct pulsations.}

\articlefiguretwo{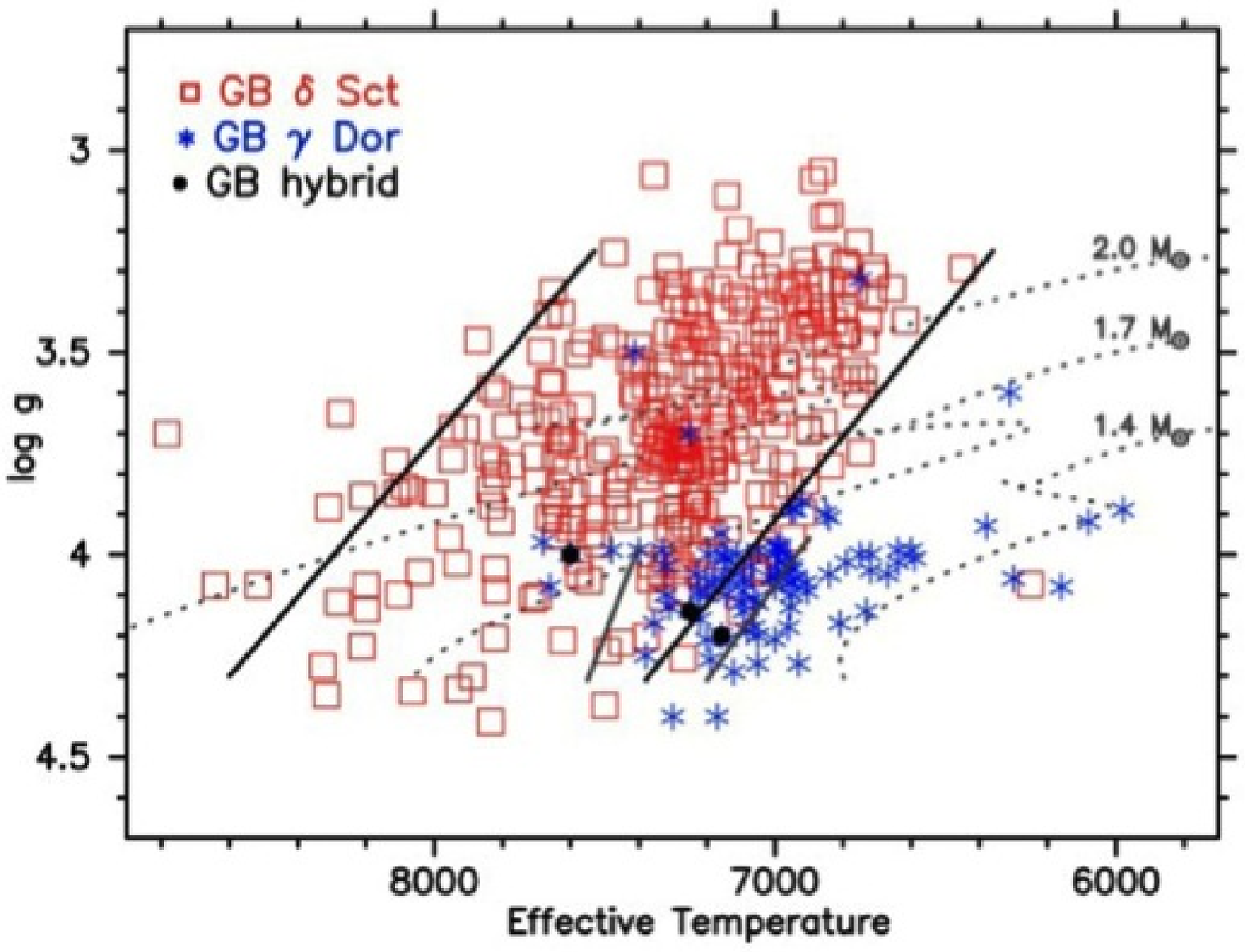}{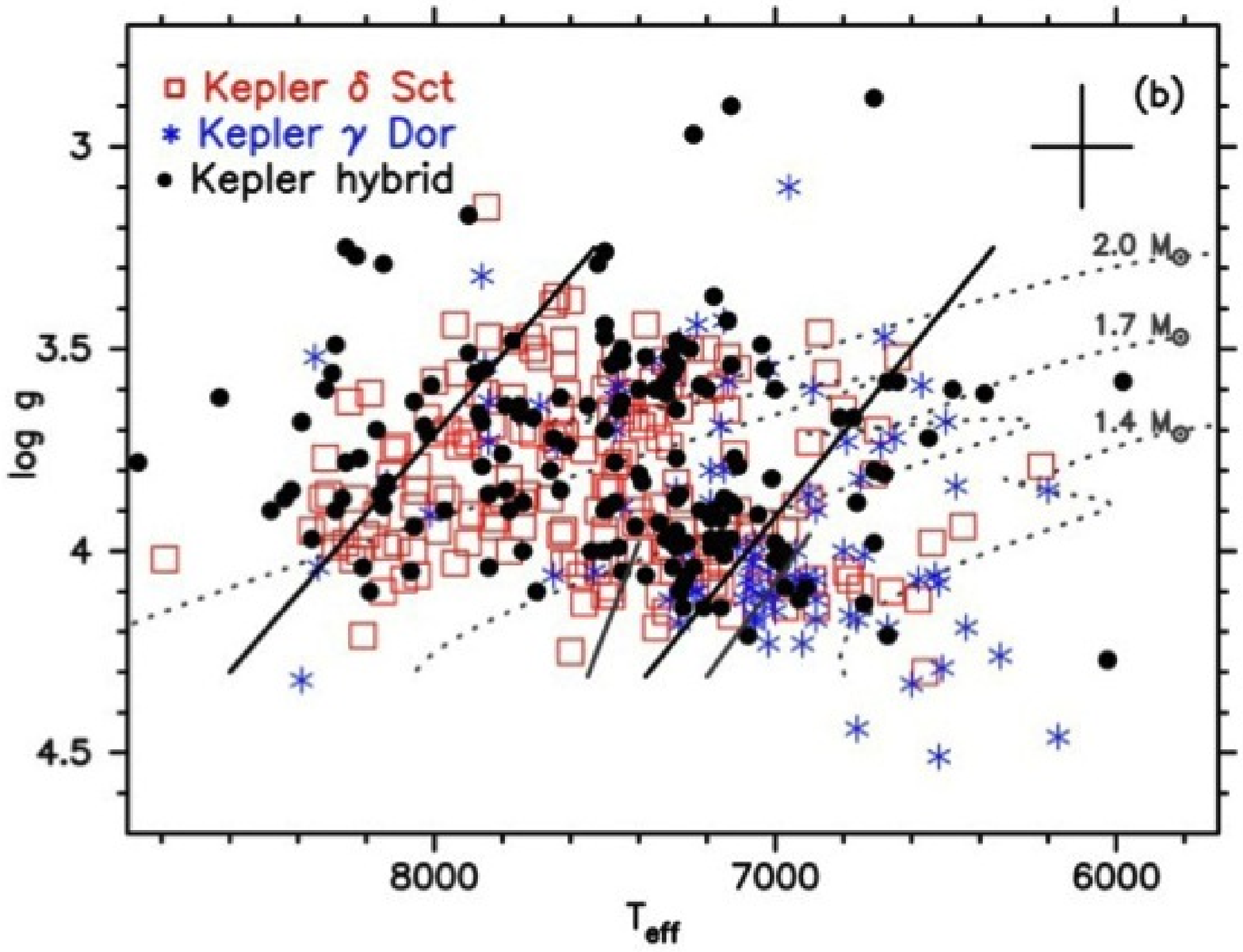}{IS}{Distribution of $\gamma$ Dor and $\delta$ Sct stars and instability strips determined from ground-based observations pre-{\it Kepler} (left), and distribution of $\gamma$ Dor and $\delta$ Sct stars discovered by {\it Kepler} observations (right).  The solid lines mark the boundaries of the $\gamma$ Dor and $\delta$ Sct instability regions determined from ground-based observations.  See \citet{2011A&A...534A.125U} for additional discussion.    Reproduced with permission from Astronomy \& Astrophysics, \copyright~ESO.}

$\delta$ Sct pulsations are driven by the $\kappa$ effect in the He+  ionization region around 50,000 K (Figure~\ref{Driving}, right), while $\gamma$ Dor {\it g}-mode pulsations are  predicted to be driven by the convective blocking mechanism at the base of the envelope convection zone around 300,000 K (Figure~\ref{Driving}, left).  Hybrid stars with pulsations of both types are predicted by theory only in a small overlapping region where the convection zone is deep enough to drive the {\it g} modes, but the convection is not too efficient to destroy the $\kappa$ effect \citep{2005A&A...435..927D}.  This picture seemed consistent with observations of known $\gamma$ Dor and $\delta$ Sct stars pre-{\it Kepler} (Figure~\ref{IS}, left), but {\it Kepler} observations of A and F stars revealed stars of both types and hybrids outside of their expected instability regions \citep{2011A&A...534A.125U,2010ApJ...713L.192G,2011MNRAS.415.3531B} (Figure~\ref{IS}, right).  The {\it Kepler} data motivate the need to investigate some additional driving mechanisms for both types of pulsations, e.g., stochastic excitation as in solar-like oscillations and red giants \citep{2010arXiv1003.4427C}; convective driving as in DA white dwarfs \citep{1999ApJ...511..904G}; Fe-concentration from diffusive settling and levitation \citep{2000A&A...360..603T,2009ApJ...704.1262T} as proposed to explain the pulsations of subdwarf B stars {\citep{2003ApJ...597..518F}; or a modified $\kappa$ effect involving both the Fe and He ionization opacity bumps \citep{1996DSSN...10...13G,2000ASPC..203..447L}.

\subsection{Models and observations of $\theta$ Cyg}

{\it Kepler}'s brightest target, the F4V main-sequence star $\theta$ Cyg was expected to be a $\gamma$ Dor star, but was discovered by {\it Kepler} observations to be a solar-like oscillator (see \citet{2011arXiv1110.2120G} and references therein).  Finding hybrid $\gamma$ Dor/ solar-like oscillators would provide additional constraints on the structure of these stars that may shed light on many of the problems outlined in this review.

Using the observed solar-like oscillation spectrum as a constraint, models were constructed for $\theta$ Cyg by several teams using different initial abundances, modeling assumptions, and evolution codes. For the preferred mode identification scenario, the convection zone depths of the models vary between 320,000 K (nearly optimum for driving $\gamma$ Dor pulsations), to 540,000 K (too deep for $\gamma$ Dor pulsations).  See Guzik et al. (2014, in preparation) for additional detail.

\citet{2011arXiv1110.2120G} constructed two models of $\theta$ Cyg that bracketed the T$_{\rm eff}$ and luminosity determined from ground-based observations, and the large separation of solar oscillation frequencies.  The 1.38 M$_{\odot}$ model has initial element mass fraction Z = 0.017, close to that of the GS98 solar abundance, while the 1.29 M$_{\odot}$ model has Z= 0.013, closer to the AGS05 abundance.  Figure~\ref{13CygHRD} shows the evolution tracks of the two models along with the $\theta$ Cyg constraints.  For the 1.38 M$_{\odot}$ model, with temperature at the convection zone base of 495,000 K, our nonadiabatic pulsation code predicts one unstable $\ell$ = 1 {\it g} mode with period 0.77 days; for the 1.29 M$_{\odot}$ model with temperature at the convection zone base 457,000 K, eight $\ell$ = 1 {\it g} modes  with periods of 0.58 to 0.95 days, and nine $\ell$ = 2 {\it g} modes with periods 0.37 to 0.55 days are predicted.  These models did not include diffusive settling;  diffusion is predicted to rapidly deplete the surface of these stars of He and most of their metals, so the fact that we see metals in these stars indicates that some process is inhibiting diffusion entirely or in part.  No {\it g} modes have been identified in the {\it Kepler} $\theta$ Cyg data to date; however, detection (or non-detection) of  {\it g} modes in $\theta$ Cyg or possibly other F stars in conjunction with additional observational constraints may help to provide a consistent modeling picture for F and G stars that would also provide constraints on diffusive settling and convection modeling, test $\gamma$ Dor pulsation driving, and perhaps provide clues to the solar abundance problem.

\articlefigure[width = 7.5 cm]{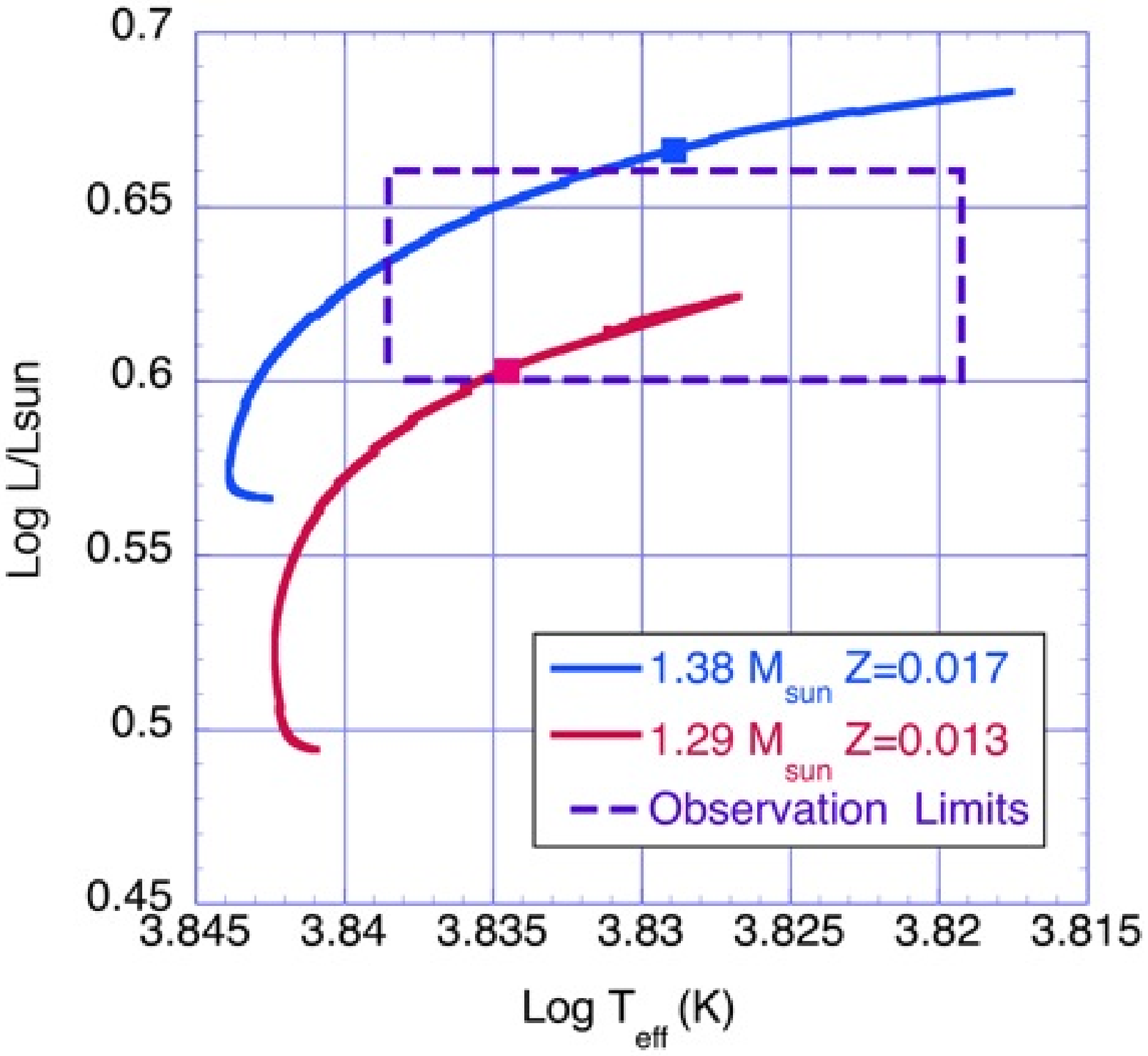}{13CygHRD}{Evolution tracks for two models of $\theta$ Cyg that bracket the observed solar-like oscillation large frequency separation, with Z abundance near the GN93 and AGS05 solar mixtures. The 1.29 M$_{\odot}$ model is predicted to show $\gamma$ Dor {\it g}-mode pulsations.}

\section{Concluding remarks}
Data from both ground and space observations are revealing new challenges to stellar model physics and providing new opportunities to test models.  We should be vigilant to look for a new level of detail of disagreement with data--other stars will help us to better understand the problems with the sun as they may be magnified in a different environment.  

\acknowledgements Author thanks the organizers for the opportunity to present this review, and the NASA {\it Kepler} Guest Observer program for funding to attend this workshop.  JAG acknowledges input from many colleagues, in particular S. Wood, K. Mussack-Tamashiro, P. Bradley, A.N. Cox, D. Pesnell, K. Uytterhoeven, J. Silk, and L.S. Watson.


\bibliography{GuzikNSO27Jan2014}

\end{document}